# Planar Edgeless Silicon Detectors for the TOTEM Experiment

G. Ruggiero, E. Alagoz, V. Avati, V. Bassetti, V. Berardi, V. Bergholm, V. Boccone, M. Bozzo, A. Buzzo,
M. G. Catanesi, R. Cereseto, S. Cuneo, M. Deile, R. De Oliveira, K. Eggert, N. Egorov, I. Eremin, F. Ferro, J. Hasi,
F. Haug, R. Herzog, P. Jarron, J. Kalliopuska, A. Kiiskinen, K. Kurvinen, A. Kok, W. Kundrát, R. Lauhakangas,
M. Lokajíček, D. Macina, M. Macrí, T. Mäki, S. Minutoli, L. Mirabito, A. Morelli, P. Musico, M. Negri,
H. Niewiadomski, E. Noschis, F. Oljemark, R. Orava, M. Oriunno, K. Österberg, V. G. Palmieri, R. Puppo,
E. Radicioni, R. Rudischer, H. Saarikko, G. Sanguinetti, A. Santroni, P. Siegrist, A. Sidorov, G. Sette, J. Smotlacha,
W. Snoeys, S. Tapprogge, A. Toppinen, A. Verdier, S. Watts, and E. Wobst

*Abstract*—Silicon detectors for the Roman Pots of the the large hadron collider TOTEM experiment aim for full sensitivity at the edge where a terminating structure is required for electrical stability. This work provides an innovative approach reducing the conventional width of the terminating structure to less than 100 $\mu$m, still using standard planar fabrication technology. The objective of this new development is to decouple the electric behavior of the surface from the sensitive volume within a few tens of micrometers. The explanation of the basic principle of this new approach together with the experimental confirmation via electric measurements and beam test are presented in this paper, demonstrating that silicon detectors with this new terminating structure are fully operational and efficient to under 60 $\mu$m from the die cut.

## I. Introduction

THE TOTEM experiment will detect the large hadron collider (LHC) leading protons at special beam pipe insertions called Roman Pots. The detectors inserted in the roman

G. Ruggiero, E. Alagoz, M. Deile, R. De Oliveira, K. Eggert, F. Haug, P. Jarron, D. Macina, H. Niewiadomski, E. Noschis, M. Oriunno, P. Siegrist, W. Snoeys, and A. Verdier are with the European Organization for Nuclear Research (CERN), Physics Department, 231211 Geneva, Switzerland (e-mail: gennaro.ruggiero@cern.ch).

V. Avati, V. Bergholm J. Kalliopuska, A. Kiiskinen, K. Kurvinen, R. Lauhakangas, T. Mäki, F. Oljemark, R. Orava, K. Österberg, V. G. Palmieri, H. Saarikko, S. Tapprogge, and A. Toppinen are with the High Energy Physics Division, Department of Physical Sciences, University of Helsinki and Helsinki Institute of Physics, 02015 Helsinki, Finland.

V. Bassetti, V. Boccone, M. Bozzo, A. Buzzo, R. Cereseto, S. Cuneo, F. Ferro, M. Macrí, S. Minutoli, A. Morelli, P. Musico, M. Negri, R. Puppo, A. Santroni, and G. Sette are with the Istituto Nazionale di Fisica Nucleare (INFN)—Sezione di Genova and Universitá di Genova, I-16146 Genoa, Italy.

V. Berardi, M.G. Catanesi, E. Radicioni are with INFN Sez. di Bari and Politecnico di Bari, Bari, Italy.

N. Egorov and A. Sidorov are with the Research Institute of Material Science and Technology, Zelenograd, Moscow 124498, Russia.

I. Eremin is with the Russian Academy of Sciences, Megaimpulse/Ioffe Physico-Technical Institute, St. Petersburg 190121, Russia.

J. Hasi, A. Kok, and S. Watts are with the Electric and Computer Engineering Department, Brunel University, UB8 3PH Uxbridge, U.K.

R. Herzog, R. Rudischer and E. Wobst are with the ILK, Institut für Luft und Kältetechnik, 01069 Dresden, Germany.

W. Kundrát, M. Lokajíček and J. Smotlacha are with the Institute of Physics, Academy of Sciences of the Czech Republic, 162 53 Praha, Czech Republic.

L. Mirabito is with the Institute de Physique Nucleaire de Lyon, 69622 Lyon, France.

G. Sanguinetti is with the Istituto Nazionale di Fisica Nucleare (INFN)—Sezione di Pisa, 56100 Pisa, Italy.

pots have to fulfil stringent requirements set by the machine and the TOTEM experiment [1]. During operation the detector edge is positioned at a distance of less than 1 mm from the axis of the high intensity proton beam where a 200-$\mu$m window separates the detectors from the primary beam vacuum. For optimal performance, the detector has to approach the $10\sigma$ envelope of the beam as closely as possible. Consequently, the detectors should be active up to their physical edge. It is our aim that the active volume should be within 50 $\mu$m of the edge.

In general, planar silicon detectors have a wide (0.5–1 mm) insensitive border region around the sensitive area. This insensitive region is occupied by a sequence of guard rings which controls the potential distribution between the detector's sensitive area and the die cut to minimize the electrical field and, thus, the surface leakage current [2], [3]. In this paper, a new approach to reducing this region will be described.

## II. Current Terminating Structure

### A. Conception

After separating the dice contained in a wafer with a diamond saw, each die presents a high density of lattice defects, dangling bonds and disordered regions. It is known that a significant fraction of these defects is electrically active, i.e., they produce energy levels in the silicon forbidden gap [4], [5]. Given the wide variety of the defects, their energy levels can be considered almost continuously distributed between the valence and conduction bands. Their high concentration is responsible for a high conductivity of the cut surface producing an effective screening of the electric field in the layers adjacent to the chip cut. On the other hand in the presence of an oxidizing atmosphere a naturally grown layer of $SiO_2$ appears on the cut surface reducing its conductivity. All this means that the final properties of the cut surface are not well determined.

The need for independence from this wide range of possible boundary conditions as close as possible to the edge has driven to the development of a new approach for terminating structures in radiation silicon detectors. The basic idea is to apply the full detector bias across the detector chip cut and collect the resulting leakage current on an outer ring, which surrounds the active area and which is biased at the same potential as the detecting strips (see Fig. 1). This ring is separated from the detector biasing electrode (the strips are biased by means of a punch-through structure between this biasing electrode and the strips). Separating and biasing these two rings at the same potential strongly reduces the influence of the current generated at the detector edge on the active detector volume. In contrast with other ring structures which provide voltage termination, this structure terminates the current and, therefore, we have called it a "current terminating structure" (CTS).

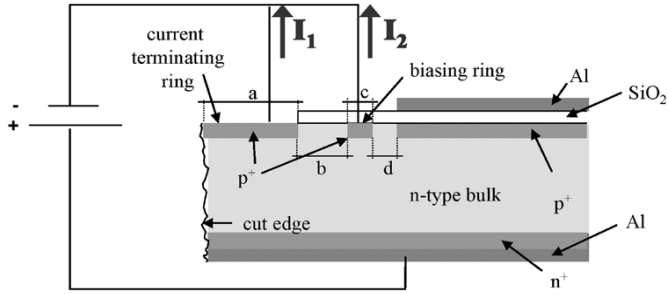

Fig. 1. Cross section of a silicon detector with a CTS in the plane parallel to the strips and its biasing scheme. In the drawing are also shown the characteristic widths of the CTS, i.e., (a) the width of the CTR and (c) of the biasing ring, (b) their distance and (d) the distance between the BR and the end of the strips.

TABLE I
CHARACTERISTIC WIDTHS OF THE FOUR TOPOLOGIES OF DETECTORS WITH CURRENT TERMINATING STRUCTURE

| Detector Type | a (μm) | b (μm) | c (μm) | d (μm) |
|---|---|---|---|---|
| A | 20 | 10 | 5 | 6 |
| B | 20 | 20 | 5 | 6 |
| C | 40 | 10 | 5 | 6 |
| F | 60 | 20 | 5 | 6 |

*B. Device Description*

These first silicon detectors produced with the CTS have been developed in a joint effort between the TOTEM group at CERN and the Megaimpulse, a spin-off company from the Ioffe PT Institute in St. Petersburg, Russia. A simplified cross section of detectors with the CTSs at the sensitive edge together with the biasing scheme is presented in Fig. 1. These devices were microstrip detectors of dimension 1 cm × 1 cm with pitches of 50 and 100 μm processed on a very high resistivity N-type silicon wafer (∼8 kΩ cm), 350 μm thick. All of them had CTS with the current terminating ring (CTR) surrounding the whole sample and AC coupled strips biased through the bias ring (BR) which is placed between the CTR and the sensitive bulk. These detectors were produced with four different topologies of CTS. They differed in the width of the CTR and its distance from the biasing ring. The characteristic widths of the four topologies as shown in the Fig. 1 are summarized in Table I. For all the detectors the distance from the end of the strips to the die's cut ranged from 41 μm to 91 μm.

The distance between the CTR and the biasing ring on the other three sides of the dice was of the order of hundreds of micrometers and equal for all the different topologies.

The picture of a corner of a sample of type "B" with strip pitch of 50 μm is shown in Fig. 2. The picture shows the features of the CTS such as the thinning of the biasing ring at the sensitive edge to 5 μm.

*C. Thermoelectric Characterization*

To study the fractions of the surface current flowing in the sensitive volume of the detector with CTS, the produced sam-ples were tested on a sample holder with BR and CTR both bonded to allow current/voltage (I–V) and current/temperature (I–T) measurements.

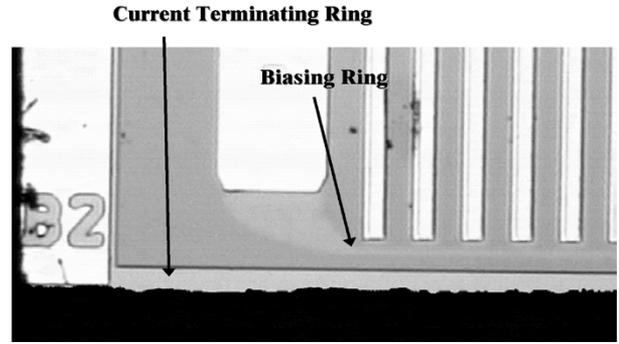

Fig. 2. Detail of the edge of a microstrip silicon detector with CTS. With this type of terminating structure the cut of the die can be even just 40 μm away from the end of the strips.

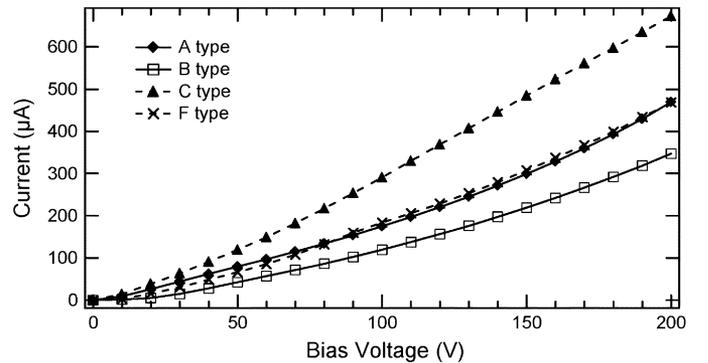

Fig. 3. I–V characteristics measured at the CTR for the different topologies.

The I–V characteristic dependence measured at the CTR for the different topologies is shown in Fig. 3.

The current measured at the CTR is strongly dominated by the current generated at the surface. If this current flows even partially in the active region, it will make the operation of the detector impossible.

The variation of this current for the different samples does not seem to be correlated with the changes in their ring structures at the edge, but seems to be more an effect of the differences that can arise at the surface after the cut.

In any case, given the cutting technique (diamond saw) some variations are expected.

This is not the case for the current flowing in the BR as shown in Fig. 4 for the same samples: the current on the BR is less by up to four orders of magnitude compared to the one flowing in the CTR.

The low current flowing in the biasing ring confirms the validity of the current termination approach: the sensitive bulk, even if extends to a few tens of micrometers from the cut edge is free of the large current flowing at the surface. Moreover the difference between the different topologies investigated seems to be negligible.

This set of detectors depletes fully at a reverse bias of around 20 V and was shown to be stable for biases higher than 200 V.

In order to study the nature of the bulk and the surface currents their behavior with different temperature was also measured. A typical temperature dependence of these two currents is shown in Fig. 5, for a reverse bias of 100 V.

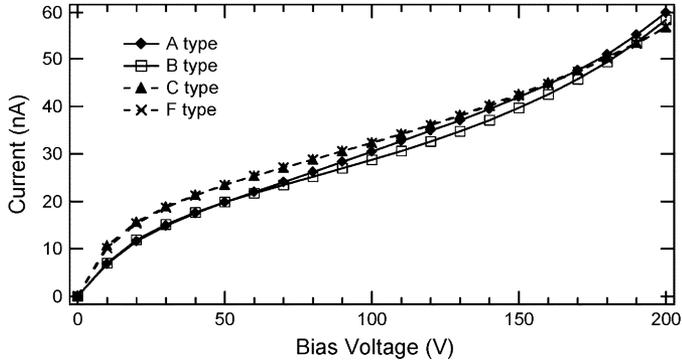

Fig. 4. I–V characteristics measured at the biasing ring for the different topologies. The current reduction is evident and illustrates the effectiveness of grounding the outer ring.

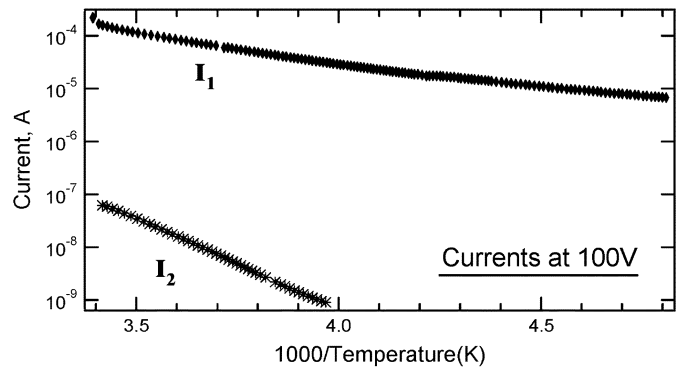

Fig. 5. Arrhenius plot for the current flowing through the CTR ($I_1$) and the biasing ring ($I_2$) of the detector of type A. The substantial difference of four orders of magnitude at room temperature increases at lower temperatures.

The current collected by the BR shows an exponential increase with the temperature. This behavior is consistent with current generation via energy levels in the middle of the silicon band gap. The origin of these levels could be related to the production technology or to defects generated in the external part of the sensor by mechanical stress at the cut of the die. On the other hand, the current flowing through the outer ring increases with the temperature but not exponentially and less steeply than the bulk current. Nevertheless, it is worth stressing that these data still confirm the basic idea of the CTS and the decoupling of the bulk current from the surface current (a difference of four orders of magnitude at room temperature and even higher at lower temperatures) and show that a further reduction of the surface current with temperature is possible: in order to halve this current component, it is sufficient to cool down to $-20\,°C$.

## III. TEST BEAM

### A. Experimental Setup

Silicon detectors with the CTS of type A and B (see Table I) have been tested in September 2003 with a muon beam in the X5 area at CERN. For the test beam a special board hosting detectors and front-end electronics was produced [1]. On one side of each board, a pair of test detectors (TDs) of the same topology was mounted with the cut edges facing each other and parallel (see Fig. 6). The detectors were aligned under a microscope and the mechanical distance between the detectors was measured within a precision better than $10\,\mu m$. A reference detector (RD) was mounted on the other side of the boards with strip direction perpendicular to the ones of the TDs, i.e., parallel to the sensitive edges of the two TDs. Thus, due to the high spatial resolution of the RD (with $50\text{-}\mu m$ of strip pitch), the insensitive distance between the two TDs can be measured precisely and can be compared with the mechanical distance enabling a precise determination of the efficiency drop at the edges of the TDs.

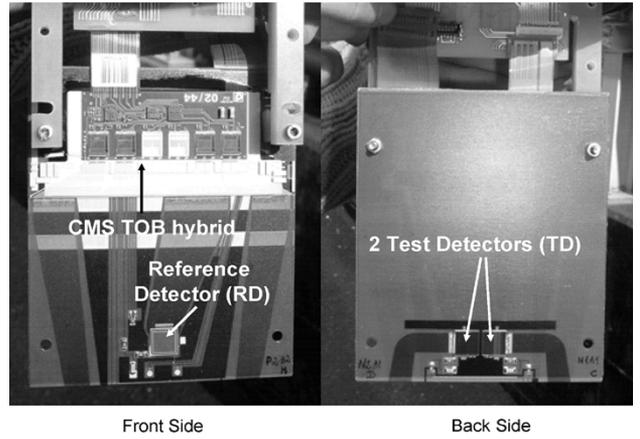

Fig. 6. Picture of the front and back side of the board developed for the test beam, hosting both test and RDs with the readout electronics.

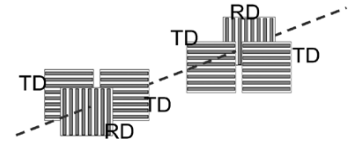

Fig. 7. Arrangement of the TDs and RDs with respect to the beam axis (dashed line).

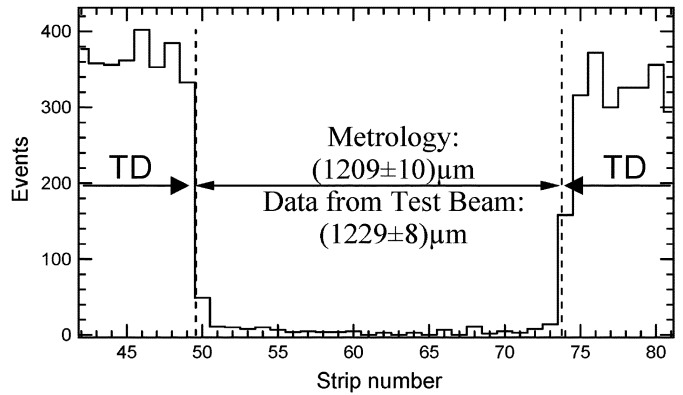

Fig. 8. Distribution of hits in the RD, in coincidence with hits in the two TDs, compared to the beginning of the sensitive area of the two TDs (dashed line).

The silicon devices were coupled with the electronics foreseen for the Roman Pot detectors in the TOTEM experiment, i.e., the APV25 chip [6], developed within the CMS collaboration [7] for the readout of the tracker. All the detectors were operated over-depleted, with a bias voltage above 110 V. The measurements were performed at room temperature. The detectors were triggered by a $10 \times 10\,mm^2$ scintillation counter, placed 2 m away from the detectors upstream on the beam line. In the experiment two boards

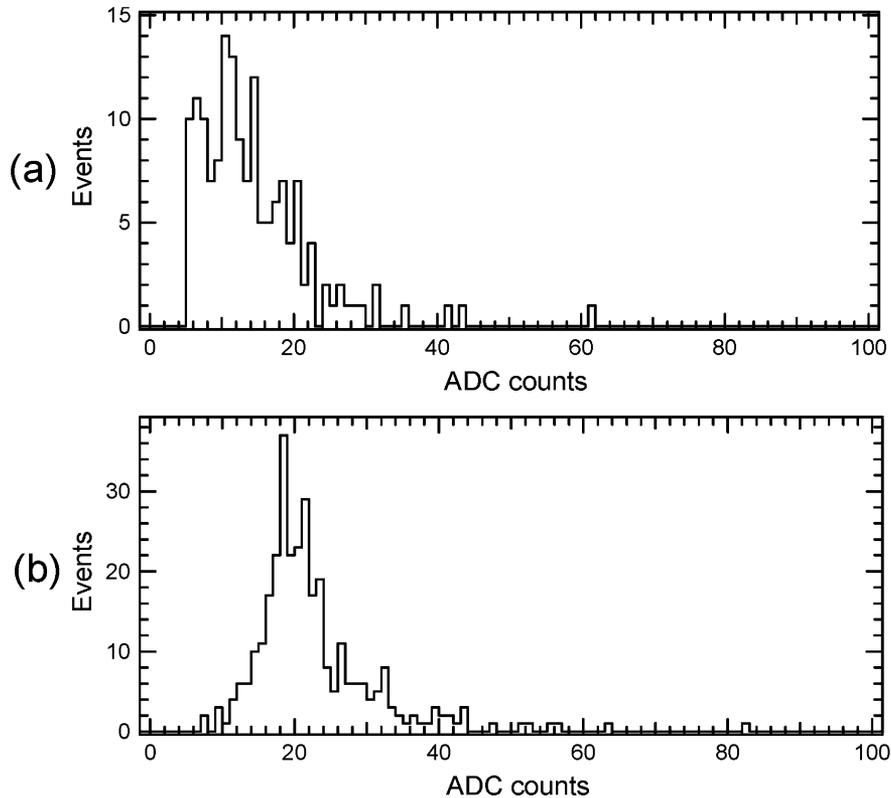

Fig. 9. (a) S/N distribution of the TD for hits at the end of the strips as recorded by the RD in the strip number 74 of Fig. 8 (a) and 50 $\mu$m away, corresponding to the strip number 75 (b).

were placed one against the other along the beam axis as shown in Fig. 7.

*B. Analysis and Results*

Tracks were defined by one hit in each RD in coincidence either with the left or with the right TDs.

The distribution of the hits in one RD which are in coincidence with a hit in one of the two TDs of type A mounted on the other side is plotted in Fig. 8. The end of the strips at the cut edge of each detector was measured with micrometric precision ($\sim$10 $\mu$m) with respect to the 50-$\mu$m strip pitch of the corresponding RD. The dashed lines in the plot give the position of the strip ends that are 40 $\mu$m away from the cut.

In principle, with good statistics, these edges can be determined with high precision from the distributions in Fig. 8. We estimate a combined statistical and systematic error of 20 $\mu$m. Since the strips start 40 $\mu$m away from the physical edge the detectors exhibit an insensitive edge region of maximum 60 $\mu$m. The results of the test on detectors of type-B are similar and they are not discussed here.

The signal-to-noise (S/N) performance of the TDs, as function of the x-position recorded in the RD shows a constant value around 22 until 50 $\mu$m away from the strips end (Fig. 9). This suggests full efficiency up to this position. However, the S/N distributions at the edges show a slight decrease in the pulse height, indicating a small loss in efficiency.

## IV. CONCLUSION

Detectors with a CTS which allows a very narrow insensitive region near the die cut were successfully tested. They showed an excellent and stable performance even at room temperature with an insensitive border of less than 60 $\mu$m. In conclusion they fully meet the experimental requirements for their use in the Roman Pots and are an excellent candidate for the detection of leading protons in LHC. Moreover, to the best of our knowledge, with this development we have produced radiation silicon detectors fabricated with standard planar technology and operated at room temperature with the smallest insensitive region at the edge.

## ACKNOWLEDGMENT

The authors are grateful to C. Da Viá for useful discussions, to A. Honma for useful discussions and for providing the CMS TOB hybrids used in the test beam, to I. McGill for helping us in setting up, mounting and bonding the modules, and to S. Roe for revising the document.